\documentclass[onecolumn,floatfix,aps,superscriptaddress]{revtex4}
\usepackage{graphicx}
\usepackage{amsmath}
\usepackage[colorlinks=true, citecolor=blue, urlcolor=blue, linkcolor = blue]{hyperref}
\usepackage{amssymb}
\usepackage{bm}
\usepackage{color}
\usepackage{bookmark}
\usepackage{tabularx}
\usepackage{mathtools}
\usepackage{microtype}
\usepackage{cancel}
\usepackage{relsize}

\usepackage{xcolor}
\begin{document}
\title{Bending of pinned dust ion acoustic solitary waves in presence of charged space debris}
\author{S. P. Acharya
\footnote{Electronic mail: siba.acharya@saha.ac.in and siba.acharya39@gmail.com}}
\affiliation{Saha Institute of Nuclear Physics, a Constituent Institute of Homi Bhabha National Institute (HBNI), 1/AF Bidhannagar, Kolkata-700064 (India)}
\author{A. Mukherjee
\footnote{Electronic mail: abhikmukherjeesinp15@gmail.com}}
\affiliation{Physics and Applied Mathematics Unit, Indian Statistical Institute, Kolkata, India}
\author{M. S. Janaki
\footnote{Electronic mail: ms.janaki@saha.ac.in}}
\affiliation{Saha Institute of Nuclear Physics, a Constituent Institute of Homi Bhabha National Institute (HBNI), 1/AF Bidhannagar, Kolkata-700064 (India)}
\begin{abstract}
We consider a low temperature plasma environment in the Low Earth Orbital (LEO) region in
presence of charged space debris particles. The dynamics of (2+1) dimensional nonlinear dust
ion acoustic waves with weak transverse perturbation, generated in the system is found to be governed by a forced
Kadomtsev-Petviashvili (KP) equation, where the forcing term depends on charged space
debris function. The bending phenomena of some exact dust ion acoustic solitary wave
solutions in $x-t$ and $x-y$ plane  are shown; that are resulted from consideration of different types of possible localized
debris functions. A family of exact pinned accelerated solitary wave solutions has been 
obtained where the velocity changes over time but the amplitude remains constant. The shape
of debris function also changes during its propagation. Also, a special exact solitary wave
solution has been derived for the dust ion acoustic wave; that gets curved in spatial dimensions with the
curvature depending upon nature of forcing debris function. Such intricate solitary wave
solutions may be useful in modelling real experimental data.
\end{abstract}
\maketitle

\section{Introduction} \label{intro}
Upsurge in the research on space plasma physics has began and achieved importance from
middle of last century \cite{Spaceplasma}.
As a result, near-Earth space
has become a virtual laboratory to study different physical properties of astronomical plasma
system. In this context, the research on dynamics of
space debris objects has also increased drastically \cite{Klinkrad, Sampaio, Kulikov}.

Space
debris objects \cite{Klinkrad} include dead satellites, meteoroids, destroyed spacecrafts and
other inactive materials resulting from many natural phenomena etc.; which are being levitated
in extraterrestrial regions especially in near-Earth space. The space debris objects are
substantially found in the Low Earth Orbital (LEO) and Geosynchronous Earth Orbital (GEO)
regions \cite{Sampaio}. Also, their number is continuously increasing nowadays due to
various artificial space missions which result in dead satellites, destroyed spacecrafts etc. The
debris particles become charged in a plasma medium because of different mechanisms such as
photo-emission, electron and ion collection, secondary electron emission \cite{Horanyi} etc.
These charged debris particles of varying sizes ( as small as microns to as big as centimeters)
\cite{Sen, NASA} move with different velocities causing significant harm to running spacecrafts.
Therefore, to avoid these deteriorating effects, active debris removal (ADR) has become a
challenging problem in twenty-first century. Some indirect detection techniques for space debris
objects have also been developed by different authors \cite{Kulikov, Sen, AcharyaHall}.
A new detection technique for charged debris objects in the LEO region was
proposed by Sen et al \cite{Sen} making use of the interaction of nonlinear plasma
waves with charged space debris. They considered the dynamics of nonlinear
ion acoustic waves in (1+1) dimensions and derived a forced KdV equation
where the forcing term depends on the debris function. Considering a Gaussian debris function,
 they have predicted  ``precursor solitons'', that are emitted
periodically, through numerical computation. Such precursors  can be detected by appropriate sensors to give an indirect evidence of the
existence of charged space debris objects. Numerical techniques \cite{Sen1} have confirmed
the existence of the pinned and precursor solitons even in the large amplitude
generalization of the problem. Molecular dynamic simulations for a charged
object moving at a supersonic speed in a strongly coupled dusty plasma
have also shown \cite{SenMD} the existence of precursor solitonic pulses and
dispersive shock waves. Experimental investigations \cite{Sen2} carried out in
dusty plasmas have corroborated the numerical findings and also observed \cite{Sen3}
modifications in the propagation characteristics of precursor solitons due
to the different shapes and sizes of the object over which the dust fluid
flows.

In the ionospheric plasma region, the presence of dust particles having different
charges and polarities cannot be ignored. These dust particles are weakly coupled as in most
space and astrophysical dusty plasmas. 
Shukla and Silint \cite{Shukla1}, first theoretically
investigated the existence of low frequency dust ion acoustic waves in plasmas.
The presence of static charged dust particles in a plasma significantly affects the physical properties
of Dust Ion-Acoustic Waves (DIAWs).
Numerous theoretical works \cite{ShuklaMamunBook}-\cite{Bharuthram} have
been reported
on the studies of linear low-frequency modes as well as the associated nonlinear coherent
structures such as solitons, shocks and vortices in different conditions of space and laboratory
dusty plasmas. The excitation of  linear and nonlinear perturbations in the upstream and
downstream regions due to a charged moving object in such dusty plasmas has many practical
implications in the context of space debris detection that deserve detailed investigations.
Generally, the sub-centimetre sized debris particles in the ionospheric plasma are currently
undetectable through optical means \cite{Sen, Truitt}. The investigation of nonlinear ion acoustic
waves in presence of charged debris by different authors \cite{Sen, Truitt, TruittKP, Mukherjee}
has primarily intended to detect these sub-centimetre sized debris particles. As discussed
earlier, there are computational and experimental works \cite{SenMD, Sen2, Sen3} in dusty
plasmas with consideration of moving charged objects, reporting observation of wakes and
precursor solitons. The observations have been shown to be consistent with theoretical
predictions based on forced KdV equation that has been derived for a charged object moving in
a plasma medium.

The forcing debris function mainly depends upon the size, shape, velocity, surface potential of
the debris, Debye shielding effects and distance of the point of observation from debris
\cite{Truitt, Truittdamp}. The mathematical form of this debris function has been derived  by
Truitt et al. \cite{Truitt} to be of the form of Gaussian function. Many authors have chosen
various kinds of forcing functions like $sech^2$ and $sech^4$ forms \cite{Sen} or sinusoidal
forms \cite{TFAli,TFChatterjee,TFZhen} to obtain solutions of the forced KdV equation. The
chaotic dynamics of nonlinear ion acoustic wave in presence of sinusoidal
source debris term in Thomas-Fermi plasmas has also been explored \cite{TFMandi}.

 In real
space plasma environment, it can be expected that both the amplitude and the velocity of the debris function may vary with
time. In \cite{Mukherjee},  a special ion acoustic solitary wave solution has been derived
in (1+1) dimensions for a specific pinned debris function, where their  velocity  vary with time.
Further, Sen. et al \cite{Sen} have shown that the $Sech^2$ form of debris function leads to
exact solutions to the forced KdV equation. The choice of a forcing term with arbitrary free
functions allows the freedom to design more interesting solutions that may bend, twist or turn
during propagation. It is possible to find certain forms of debris functions for which exact
solitary wave solutions can be obtained that have the bending features. On the other hand, the
varying velocity of the debris function turns out to be identical with that of the solitary wave
solutions; hence they are pinned to each other, and so they accelerate / decelerate together.
In the present work, some special exact (2+1) dimensional solitary wave solutions for the dust
ion acoustic wave have been derived having both accelerating and bending features. An interesting
intricate exact solution of the system is important for both experimental observations and
validation of numerical simulations.

The paper is organized in the following manner. The detailed derivation of the (2+1) dimensional
nonlinear evolution equation for the dust ion acoustic wave in presence of charged debris
object is given in section-II. The exact accelerated pinned solitary wave solutions are derived in
section-III. The exact curved solitary wave solution that can bend in $x-y$ plane depending on
the nature of debris function, is discussed in section-IV. In section-V, we discuss some
dynamical and physical properties of the obtained solutions. Concluding remarks are given in
section-VI followed by acknowledgments and bibliography.
\section{Derivation of (2+1) dimensional nonlinear evolution equation for the dust ion acoustic
waves in presence of charged space debris} \label{Evl_eqn}
We consider the low temperature plasma system in the Low Earth Orbital (LEO) region in
presence of charged space debris particles. Our aim is to find the evolution equation for the
propagating finite amplitude nonlinear dust ion acoustic waves (DIAWs) generated
in the system. As an extension to our previous problem on ion acoustic waves \cite{Mukherjee},
we consider two dimensional effects in the propagation of nonlinear dust ion acoustic waves.
The LEO region which we consider, consists of a low density plasma along with the abundance
of debris particles. The ion species is treated as a cold species, i.e. ion pressure is neglected
and electrons obey the Boltzmann distribution.
We neglect the dynamics of heavy and slow dust particles compared to the dynamics of the
other particles present in the system. Hence, only the equilibrium dust density (independent of
time) enters into the calculation through charge neutrality condition.
Following \cite{CTP, OR, Rehman}, the basic normalized nonlinear system of equations of our
system in (2+1) dimensions is given by
\begin{eqnarray}
\frac{\partial n}{\partial t}+\frac{\partial}{\partial x}(nu)+\frac{\partial}{\partial
y}(nv)=0,\label{Continuity}\\
\frac{\partial u}{\partial t}+u\frac{\partial u}{\partial x}+v\frac{\partial u}{\partial y}+\frac{\partial
\phi}{\partial x}=0,\label{Momentum_x}\\
\frac{\partial v}{\partial t}+u\frac{\partial v}{\partial x}+v\frac{\partial v}{\partial y}+\frac{\partial
\phi}{\partial y}=0,\label{Momentum_y}\\
\frac{{\partial}^2 \phi}{\partial x^2}+\frac{{\partial}^2 \phi}{\partial y^2}+n-(1-\alpha)e^{\phi}-\alpha
n_d=S(x,y,t),\label{Poisson}
\end{eqnarray}
where the following normalizations have been used:
\begin{equation}
x\longrightarrow x/{\lambda}_d; \,y\longrightarrow y/{\lambda}_d;\,t\longrightarrow
\frac{C_s}{{\lambda}_d}t; \, n\longrightarrow \frac{n}{n_{i0}};\, u\longrightarrow \frac{u}{C_s};\,
v\longrightarrow \frac{v}{C_s};\, \phi \longrightarrow \frac{e \phi }{k_BT_e},
\end{equation}
where ${\lambda}_d$ is electron Debye length, $\mathrm{C_s} = \sqrt{\frac{k_B T_e}{m_i}}$ is
ion acoustic speed, $\mathrm{k_B}$ is Boltzmann constant, $T_e$ is electron temperature,
$m_i$ is the ion mass and $n_{i0}$ is equilibrium ion density. Equations (\ref{Continuity}),
(\ref{Momentum_x}), (\ref{Momentum_y}) and (\ref{Poisson}) represent ion continuity equation,
ion momentum conservation equations in $x$ and $y$ directions, and Poisson's equation
respectively; where n, u, v, and $\phi$ denote the ion density, x and y component of ion fluid
velocity and electrostatic potential respectively. The parameter $\alpha$ in the LHS of Poisson
equation (\ref{Poisson}) is given by
\begin{equation}
\alpha = Z_d \frac{n_{d0}}{n_{i0}}, \label{Alpha}
\end{equation}
where $Z_d$ and $n_{d0}$ represent dust charge and equilibrium dust density respectively.

 The term $S(x,y,t)$ in the RHS of equation (\ref{Poisson}) represents a charge density source
arising due to a  time varying debris object having a two dimensional space dependence,
which creates a perturbation in electric potential and density of the surrounding plasma. As a
result, the plasma potential becomes modified due to debris surface potential; which is
discussed in detail in the work of Truitt et al. for one dimension \cite{Truitt, Truittdamp} and for
two dimensions \cite{TruittKP}. In their work, Truitt et al. have  described this forcing or
source debris function $S(x,y,t)$ to be of Gaussian nature in both one and two dimensions using
the work of Grimshaw et al. \cite{Grimshaw}. The amplitude of this force debris function is given
by normalized plasma potential. Sen et al. \cite{Sen} have considered localized solitary wave
forms and Gaussian form for the debris function. Taking these  developments into account,
we approximate the source debris function as
\begin{equation}
S(x,y,t)={\phi}_{pn}S_d(x,y,t), \label{debrisform}
\end{equation}
where ${\phi}_{pn}$ denotes normalized plasma potential and $S_d(x,y,t)$ represents possible
localized debris functions like Gaussian, solitary wave types etc. that move with velocities
consistent with the debris object.
In the work done by Sen et al. \cite{Sen}, ion acoustic solitary wave solutions for two specific
forms of localized forcing debris functions have been derived that also have the nature of solitary
waves. The line solitary wave solutions have constant amplitudes and velocities; which has
been generalized in  \cite{Mukherjee} by considering more realistic time dependent velocity for
both ion acoustic solitary wave and the forcing term.

The evolution equation corresponding to the nonlinear (2+1) dimensional DIAWs is derived
following the well-known reductive perturbation technique (RPT) \cite{Kraenkel}, where the
dependent variables of the system are expanded as :
\begin{eqnarray}
n=1+{\epsilon}^2n_1+{\epsilon}^4n_2+ O(\epsilon^6),\label{Expn_n}\\
u={\epsilon}^2u_1+{\epsilon}^4u_2+ O(\epsilon^6),\label{Expn_u}\\
v={\epsilon}^3v_1+{\epsilon}^5v_2 + O(\epsilon^7),\label{Expn_v}\\
\phi={\epsilon}^2{\phi}_1+{\epsilon}^4{\phi}_2 + O(\epsilon^6),\label{Expn_phi}
\end{eqnarray}
where $\epsilon$ is a small dimensionless expansion parameter characterizing the strength of
nonlinearity in the system. We consider a weak, space-time dependent localized debris function
which vanishes at infinity. After scaling we have
\begin{equation}
S(x,y,t)={\epsilon}^4f(x,y,t),\label{Expn_F}
\end{equation}
where $f(x,y,t)$ can have any spatially localized form that is consistent with the weakly coupled
charged debris dynamics in the LEO region. The stretched coordinates are introduced  as:
\begin{equation}
\xi=\epsilon(x-v_p t);\,\tau={\epsilon}^3t;\,\eta={\epsilon}^2y,\label{Scal_axis}
\end{equation}
where $v_p$ is the phase velocity of the wave in $x$ direction. Accordingly, the differential
operators are expressed in terms of the stretched variables as:
\begin{equation}
\frac{\partial}{\partial x}=\epsilon\frac{\partial}{\partial \xi};\,\frac{\partial}{\partial t}=-\epsilon
v_p\frac{\partial}{\partial \xi}+{\epsilon}^3\frac{\partial}{\partial \tau};\,\frac{\partial}{\partial
y}={\epsilon}^2\frac{\partial}{\partial \eta}.
\end{equation}
Putting these expanded and rescaled variables in equation (\ref{Continuity}) and collecting
different powers of $\epsilon$, we get
\begin{equation}
O({\epsilon}^3):\,-v_{p}\frac{\partial n_1}{\partial \xi}+\frac{\partial u_1}{\partial \xi}=0,
\label{Con_3}
\end{equation}
\begin{equation}
O({\epsilon}^5):\,-v_{p}\frac{\partial n_2}{\partial \xi}+\frac{\partial n_1}{\partial
\tau}+\frac{\partial}{\partial \xi}(u_2+n_1u_1)+\frac{\partial v_1}{\partial \eta}=0 \label{Con_5}.
\end{equation}
Similarly, using equation (\ref{Momentum_x}), we get
\begin{equation}
O({\epsilon}^3):\,-v_p\frac{\partial u_1}{\partial \xi}+\frac{\partial {\phi}_1}{\partial \xi}=0,
\label{Momt_3}
\end{equation}
\begin{equation}
O({\epsilon}^5):\,-v_{p}\frac{\partial u_2}{\partial \xi}+\frac{\partial u_1}{\partial
\tau}+u_1\frac{\partial u_1}{\partial \xi}+\frac{\partial {\phi}_2}{\partial \xi}=0. \label{Momt_5}
\end{equation}
Again, using equation (\ref{Momentum_y}), we get
\begin{equation}
O({\epsilon}^4):\,-v_p\frac{\partial v_1}{\partial \xi}+\frac{\partial {\phi}_1}{\partial \eta}=0.
\label{Momt_4}
\end{equation}
Finally, equation (\ref{Poisson}) yields:
\begin{equation}
O({\epsilon}^2):\,-(1-\alpha){\phi}_1+n_1=0, \label{Posn_2}
\end{equation}
\begin{equation}
O({\epsilon}^4):\,\frac{{\partial}^2{\phi}_1}{\partial
{\xi}^2}-(1-\alpha){\phi}_2-(1-\alpha)\frac{{{\phi}_1}^2}{2}+n_2= f. \label{Posn_4}
\end{equation}
Now equations (\ref{Con_3}), (\ref{Momt_3}), (\ref{Momt_4}), and (\ref{Posn_2}) give
\begin{equation}
v_p n_1=u_1; v_p u_1 = \phi_1; v_p^2 (1-\alpha) =1 ; \,\frac{\partial v_1}{\partial
\xi}=\frac{1}{v_p} \frac{\partial {\phi}_1}{\partial \eta}. \label{Simpl_1}
\end{equation}
Hence, the phase velocity $v_p$ is evaluated as
\begin{equation}
v_p = \pm \frac{1}{\sqrt{1 - \alpha}},\label{vp}
\end{equation}
where $\alpha$ is defined in (\ref{Alpha}) and we consider the propagation along positive $x$
axis by choosing positive sign of $v_p$ in (\ref{vp}).
From equations (\ref{Con_5}) and (\ref{Momt_5}), we get\\
\begin{eqnarray}
\frac{\partial n_2}{\partial \xi}=\frac{1}{v_p}[\frac{\partial n_1}{\partial \tau}+\frac{\partial
(u_2+n_1u_1)}{\partial \xi}+\frac{\partial v_1}{\partial \eta}], \label{n2xi}\\
\frac{\partial u_2}{\partial \xi}=\frac{1}{v_p}[\frac{\partial u_1}{\partial \tau}+u_1\frac{\partial
u_1}{\partial \xi}+\frac{\partial {\phi}_2}{\partial \xi}].\label{u2xi}
\end{eqnarray}
The relations given by equations (\ref{Simpl_1}), (\ref{n2xi}), and (\ref{u2xi}), after some
simplification, give the following nonlinear evolution equation :
\begin{equation}
{[ n_{1\tau}+ A \ n_1n_{1\xi}+ B \ n_{1\xi\xi\xi}]}_{\xi}+ C \ n_{1\eta\eta}= D \
f_{\xi\xi},\label{KP_original}
\end{equation}
where the subscripted variables denote partial derivatives. The coefficients appearing in
(\ref{KP_original}) are given as
\begin{eqnarray}
&{}A = \frac{v_p (3 - v_p^2)}{2}, \ B = \frac{v_p^3}{2}, \ C = \frac{v_p}{2}, \ D = \frac{v_p}{2}.
\end{eqnarray}
This is forced Kadomtsev-Petviashvili (KP) equation, i.e. generalization of forced KdV equation
to two dimensional space.
This is the final nonlinear evolution equation of the (2+1) dimensional nonlinear DIAWs in
presence of charged space debris. Since we have assumed $v_p$ to be positive hence,
equation (\ref{KP_original}) represents forced KP-II equation.
We normalize Eq. (\ref{KP_original}) to
\begin{equation}
[U_T + \ 6 U U_X + \ U_{XXX}]_X + U_{YY} = F_{XX}, \label{KPnorm}
\end{equation}
where, the new variables in (\ref{KPnorm}) are defined as $U = (\frac{A}{6 B}) n_1,\ \xi = X, \ T
= B \tau, \ Y = \sqrt{\frac{B}{C}} \eta, \ F = (\frac{D A}{6 B^2}) f$. Thus we can see that the
solutions of the forced KP-II equation (\ref{KP_original}) depend on the equilibrium dust
parameters i.e, on the parameter $\alpha$ as well as on the charged space debris function $f$.
In the following sections, we will find out various exact solitary wave solutions of
(\ref{KP_original}) having special bending feature. Those solitary waves can bend on $x-t$ or
$x-y$ plane depending on the functional forms of the charged debris function $f$. The variations
of the solitary wave solutions with the dust parameters will also be discussed. For mathematical
simplicity, we will concentrate on the normalized equation (\ref{KPnorm}) for further 
consideration. Finally we can transform the solutions to the old variables to explore the physical
picture.

\section{Exact pinned accelerated solitary wave solutions}
We have derived a forced KP-II equation (\ref{KPnorm}) as the evolution equation for the
nonlinear dust ion acoustic wave in presence of charged space debris particles. We know that
Eq. (\ref{KPnorm}) is in general non-integrable and not exactly solvable. But if the forcing term
obeys a definite constraint condition, the evolution equation can retain its integrability.
This theory of converting a forced system into an integrable system is called nonholonomic
deformation theory which is discussed in detail in \cite{Kundu,Kundu1}. Using this theory, we
can show that Eq. (\ref{KPnorm}) admits a special exact accelerated soliton solution under a
specific integrable condition. The velocity of the soliton varies with time whereas the amplitude
remains constant.
In this work, we will not go into the details of the integrability of the forced KP-II equation
(\ref{KPnorm}) by nonholonomic deformation theory \cite{Kundu, Kundu1}. Rather, we will
concentrate on the exact solvability of Eq. (\ref{KPnorm}).
Sen et al. \cite{Sen} have derived two exact line solitary wave solutions for their forced KdV
equation in (1+1) dimensions that have constant amplitude and velocity, for the choice of two
specific localized debris functions of $sech^2$ and $sech^4$ types. 

In \cite{TruittKP}, the (2+1) dimensional charged space debris function has been derived as 
\begin{equation}
F(X,Y,T)=N(\phi_s, r, \lambda_D)exp\lbrace -{\lambda^2_D}[{(\frac{X-V_X
T}{R_X})}^2+{(\frac{Y-V_Y T}{R_Y})}^2]\rbrace, \label{TruittDF}
\end{equation}
where $\phi_s$ denotes surface potential of charged debris, $r$ is the distance of point of
observation from debris, $\lambda_D$ represents Debye length, $V_X$ and $V_Y$ represent
the speeds of debris in $X$ and $Y$ directions respectively, and $R_X$ and $R_Y$ represent
the radii of debris in $X$ and $Y$ directions respectively. In realistic environment,  the debris object may not  have a perfect
spherical shape. A more complete and realistic picture of a
debris object can be realized from a three-dimensional Gaussian  function
in an analogous manner. This can be referred to as the three-dimensional interpretation of a
debris and can be utilized to study the evolution of dust ion acoustic waves in $(3+1)$
dimensions.
The amplitude of the forcing debris function $F$ given by equation
(\ref{TruittDF}) depends upon the surface potential of debris object, Debye length in the surrounding
plasma medium and distance of point of observation from debris. The argument of the exponential function in
equation (\ref{TruittDF}) contains the space-time coordinates along with other constants.
A few studies have been done in (1+1) dimensions by considering periodic forms of the
debris function \cite{TFAli,TFChatterjee,TFZhen}. But, several other types of localized functions can be
chosen as forcing debris functions; which need not be exactly Gaussian, but can represent
similar behavior. In an actual disturbed situation in the LEO plasma region, there may be
possibilities where the velocity of the debris may vary with time. Consequently the expression of
forcing debris function becomes modified in order to incorporate time-varying velocity, which is
also evident from equation (\ref{TruittDF}). Several analytical forms of these kinds of time
varying debris functions and the corresponding nonlinear dust ion acoustic waves can be
explored to study many interesting phenomena.
The mathematical forms of the two localized debris functions that were chosen for the
derivation of the two exact pinned solitary wave solutions in \cite{Sen}, have a definite feature. It
can be noticed that those debris functions are just constant multiples of the ion acoustic wave
and its square respectively. The solutions are mathematically interesting due to their pinned
and exact nature.
Motivated by this choice, we attempt to derive some special exact pinned solitary wave
solutions for the dust ion acoustic wave, the velocity of which varies with time depending on the
specific debris function. Also, the acceleration associated with the solitary wave is expressed
in terms of an arbitrary function that can be chosen appropriately to model the real physical
phenomena.
As stated in \cite{Truitt}, pinned solitons are produced at Low LEO region, traveling at the same
speed as the orbital debris. Since, the pinned solitons propagate with the debris, they are
not useful for on-orbit detection techniques as they would not be sensed before collision.
However, pinned solitons can be detected from the ground sensors using the same techniques
used to measure plasma density irregularities. The amplitudes of solitary wave solutions which
will be discussed in this section are constant whereas that of the debris function are variable.
Such features may give rise to new directions in the detection process.
 We know that the $sech^2 $ function looks very similar to the Gaussian function. Hence, a forcing debris function that is
estimated  to have  Gaussian shape \cite{TruittKP} can be chosen in the form of a $sech^2 $
function, and the understanding can also be generalized to higher dimensions.
This type of forcing debris function is also a two-dimensional extension of the forcing debris
functions chosen in the work of Sen et al. \cite{Sen}. 

For, the following choice of the forcing debris function as
\begin{equation}
F(X,Y,T) = a(T) \ U_m \ Sech^2[(X + \int a(T) dT - d_1 \ T+Y )/w], \label{debris2}
\end{equation}
where $U_m = \frac{(d_1 - 1)}{2}, \ w = (2/ \sqrt{(d_1 - 1)}); d_1 $ being constant and $a(T)$ is an arbitrary
function of time, an exact pinned solitary wave solution of Eq. (\ref{KPnorm}) can be obtained as
\begin{equation}
U = U_m \ Sech^2[(X + \int a(T) dT -d_1 \ T+Y )/w]. \label{solpinA1}
\end{equation}
 A comparison between equations (\ref{TruittDF}) and
(\ref{debris2}) helps in identifying various physical characteristics, i.e,  size, shape, velocity and
surface potential etc., of realistic charged debris objects in ionosphere. The velocity of the dust ion
acoustic solitary wave solution (\ref{solpinA1}) changes over time showing
accelerating/decelerating features due to the presence of the function $a(T)$ in its argument,
whereas its amplitude remains constant. Both the amplitude and velocity of the source
debris function $F$ (\ref{debris2}), change over time showing shape changing effects.
Since the analytic forms of the time-varying velocities of both $U$ and $F$ are identical, they
move together. So, they can also be regarded as ``pinned accelerated solitary waves'' as
discussed in \cite{Sen}.  

\begin{figure}
\includegraphics[width=10cm]{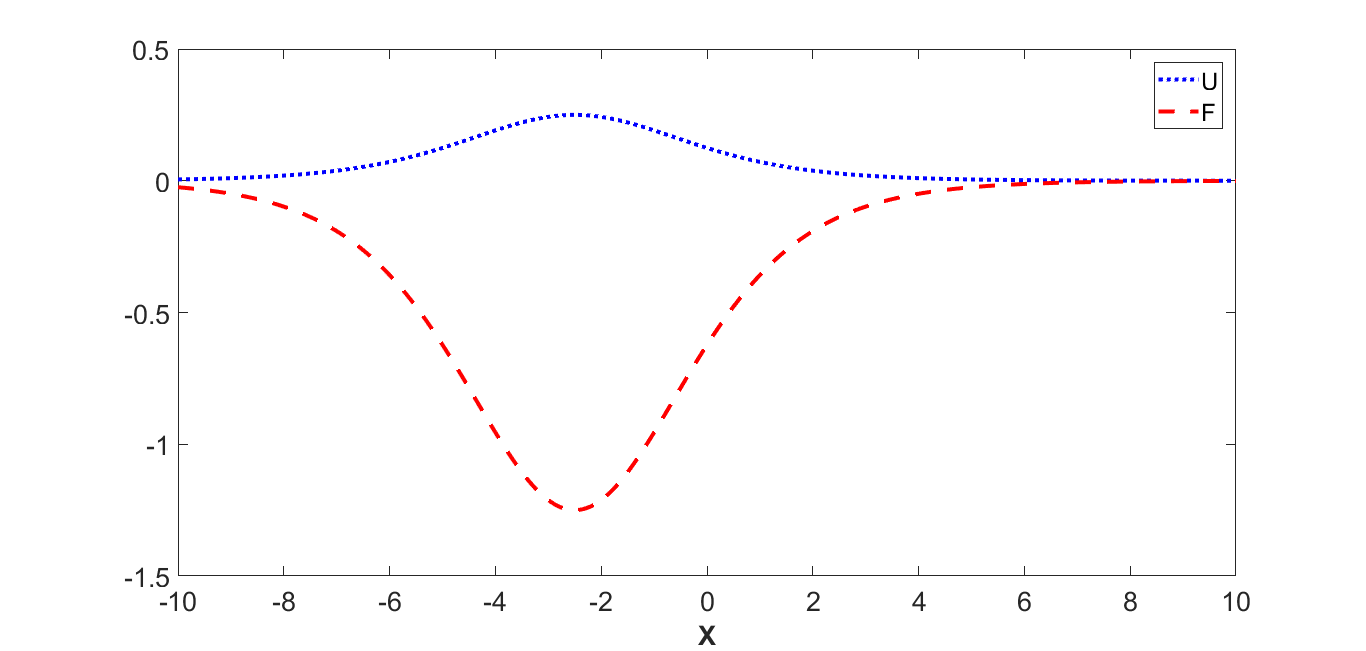}
\ \ \includegraphics[width=10cm]{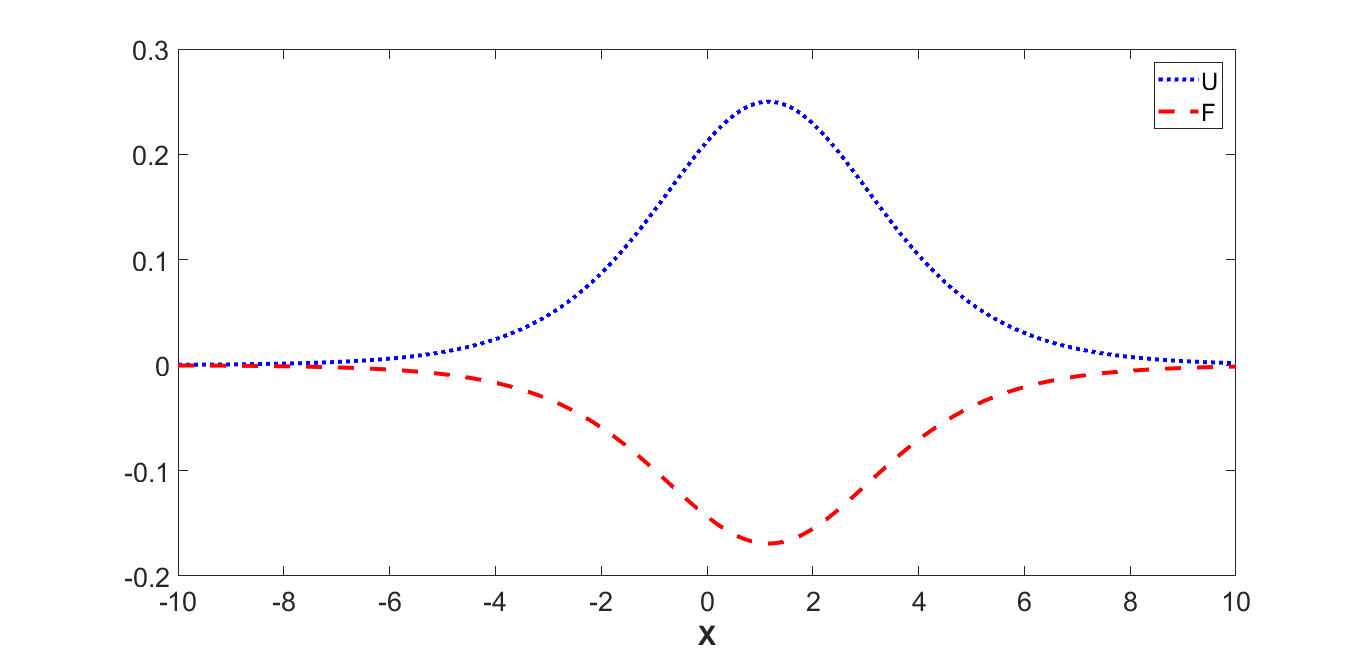}
\ \ \includegraphics[width=10cm]{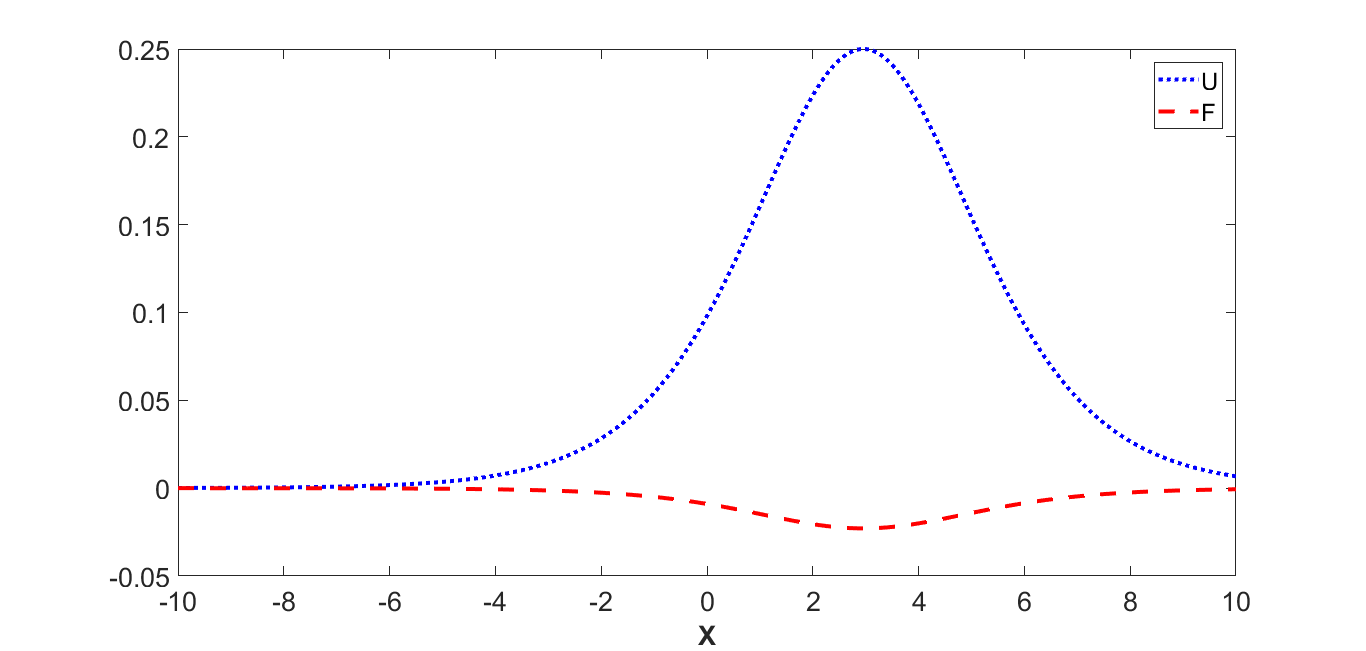}
\caption{Plots of decelerated solitary wave solution $U$ given by equation (\ref{solpinA1}) and
forcing function $F$ given by equation (\ref{debris2}) with $X$ at $Y=0$ for times $T=0$,
$T=1$ and $T=2$ respectively, for the choice of $d_1 = 1.5$ and $a(T)= - 5 e^{-2T}$. The top, middle and bottom sub-figures in the above figure indicate the natures of $U$ and $F$ at $T=0$, $T=1$ and $T=2$ respectively, with dotted blue line and dashed red line corresponding to $U$ and $F$ respectively. The plots clearly
indicate that $U$ and $F$ are pinned to each other, i.e. they move with the same velocity in
spite of having different amplitudes.} \label{Fig3}
\end{figure}

\begin{figure}
\includegraphics[width=8cm]{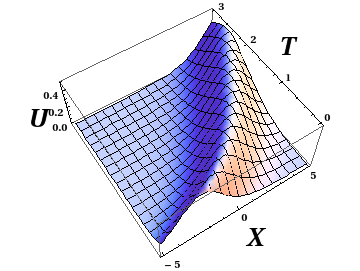}

\vspace{1cm}


\vspace{1cm}

\includegraphics[width=8
cm]{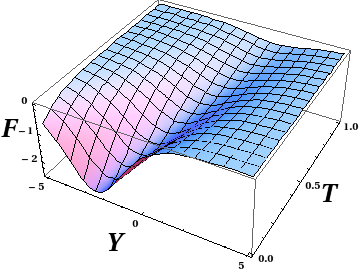}

\vspace{1cm}


\vspace{1cm}

\caption{The above figures show 3D plots of DIAW solution $U$ and debris function $F$ given
by equations (\ref{solpinA1}) and (\ref{debris2}) respectively on $X-T$ plane for $Y=0$, $a(T) = -5 e^{-2T}$ and $d_1 =1.5$. The
exponential form of $a(T)$ causes the DIAW to decelerate with time; whereas its amplitude remains constant. On the other hand, both the amplitude
and velocity of source debris function $F$ varies with time.} 
\label{Fig1}

\end{figure}

 The pinned nature of $U$ and $F$ can be seen from Figure-\ref{Fig3} for a given choice : $a(T)= - 5 e^{-(2T)}$,
where the dynamical evolution of both $U$ and $F$ have been shown by plotting the functions at
different times.
The 3D plots of $U$ and $F$ on $X-T$ plane for the same choice of $a(T)$ like before, have been shown in Figure-\ref{Fig1}.
 Though, $a(T)$ is an arbitrary function, it should be chosen according to the situation. In the plots, we have chosen the exponential form so that the solitary wave
 possesses a deceleration = $- \frac{d a(T)}{d T},$
 that can be calculated from (\ref{solpinA1}). Such kind of 
 deceleration of the solitary wave  may be relevant to the real debris problem.  For the periodic choice  $a(T) = 2 \cos{T}$, the solitary wave solution accelerates and decelerates periodically which is unsuitable for the debris problem. But it can be an interesting choice for other physical system, like the shallow water wave system. If we consider the (2+1) dimensional propagation of free surface nonlinear shallow water wave with periodic bottom boundary condition, then we can obtain solitary wave solution that decelerate and accelerate periodically. The periodic bottom boundary condition means that the bottom of the channel is porous in such a way that the downward fluid velocity is periodic function of time. Then we would get a free surface solitary  wave solution with that kind of nature.

The solutions obtained so far for one solitary wave can be generalized to a $N$ solitary wave
solution $U$ of Eq.(\ref{KPnorm}) having time varying velocity. Similarly, the $N$ debris function
$F$ that is pinned with the dust ion acoustic solitary wave also accelerates / decelerates while propagation. Using Hirota
formalism \cite{Wazwaz, BendingMukh}, the two solitary wave solution $U$ of (\ref{KPnorm})
can be evaluated as
%
\begin{align}
&{}U = 2 \ [ln G]_{XX}, \
G = 1 + \exp{(\theta_1)} +
\exp{(\theta_2)} + a_{12} \exp{(\theta_1 + \theta_2)}, \nonumber \\
&{} a_{12} = \frac{[3k_1^2 \ k_2^2 \ (k_1-k_2)^2 - (k_1 m_2 - k_2 m_1 )^2]}{[3k_1^2 \ k_2^2 \
(k_1+k_2)^2 - (k_1 m_2 - k_2 m_1 )^2]}, \ \theta_i = [ k_i X + m_i Y - \frac{k_i^4 + m_i^2}{k_i} T
+ \int a(T) dT ]
\end{align}
for the choice of the debris function
\begin{align}
&{} F = 2 a(T) \ [ln G]_{XX}, \
G = 1 + \exp{(\theta_1)} +
\exp{(\theta_2)} + a_{12} \exp{(\theta_1 + \theta_2)}, \nonumber \\
&{} a_{12} = \frac{[3k_1^2 \ k_2^2 \ (k_1-k_2)^2 - (k_1 m_2 - k_2 m_1 )^2]}{[3k_1^2 \ k_2^2 \
(k_1+k_2)^2 - (k_1 m_2 - k_2 m_1 )^2]}, \ \theta_i = [ k_i X + m_i Y - \frac{k_i^4 + m_i^2}{k_i} T
+ \int a(T) dT ],
\end{align}
where $k_i, m_i$ are constants.
Here, the amplitude of $F$ varies with time whereas that of $U$ remains constant. But both
$U$ and $F$ accelerate / decelerate with the same velocity that depends on $a(T)$. 

Generalizing  above solutions, the
accelerated $N$ solitary wave solution $U$ of (\ref{KPnorm}) can be obtained as
\begin{align}
&{} U = 2 \ [ln G]_{XX}, \nonumber \\
&{} G = \sum_{\mu=0,1,2} [\prod_{j=1}^n \ (\frac{\beta_j}{2ik_j})^{\mu_j(\mu_j-1)} \ (\beta_j
\delta_{kj} + \eta_j^{(0)})^{\mu_j(2-\mu_j)} \ \exp{(\sum_{j=1}^n \mu_j \xi_j + \sum_{1 \leq j < l
}^n \mu_j \mu_l A_{jl})} ], \nonumber \\
&{} \xi_j = k_j (\omega_j T + X + \int a(T)dT + p_j Y) + \xi_j^{(0)}, \nonumber \\
&{}\eta_l = \alpha_l T + \beta_l (X + \int a(T) dT) + \gamma_l Y + \eta_l^{(0)}, \ \gamma_j =
\beta_j p_j \nonumber \\
&{} \exp{(A_{jl})} = \frac{(k_j-k_l)^2-(1/3)(p_j-p_l)^2}{(k_j-k_l)^2-(1/3)(p_j-p_l)^2},\label{Nsol}
\end{align}
for the choice of
\begin{align}
&{} F = 2 \ a(T) \ [ln G]_{XX}, \nonumber \\
&{} G = \sum_{\mu=0,1,2} [\prod_{j=1}^n \ (\frac{\beta_j}{2ik_j})^{\mu_j(\mu_j-1)} \ (\beta_j
\delta_{kj} + \eta_j^{(0)})^{\mu_j(2-\mu_j)} \ \exp{(\sum_{j=1}^n \mu_j \xi_j + \sum_{1 \leq j < l
}^n \mu_j \mu_l A_{jl})} ], \nonumber \\
&{} \xi_j = k_j (\omega_j T + X + \int a(T)dT + p_j Y) + \xi_j^{(0)}, \nonumber \\
&{}\eta_l = \alpha_l T + \beta_l (X + \int a(T) dT) + \gamma_l Y + \eta_l^{(0)}, \ \gamma_j =
\beta_j p_j \nonumber \\
&{} \exp{(A_{jl})} = \frac{(k_j-k_l)^2-(1/3)(p_j-p_l)^2}{(k_j-k_l)^2-(1/3)(p_j-p_l)^2}, \label{Ndeb}
\end{align}
where $ k_j, \omega_j, p_j, \xi_j^{0}, \alpha_l, \beta_l, \gamma_l, \eta_l^{(0)}$ are arbitrary
real constants.
The variation of the velocities  of $U, F$ comes due to the presence of $a(T)$. Thus, we have obtained a family of special 
exact  pinned solitary wave solutions for both $F$ and $U$ having accelerating / decelerating feature, that may be useful in future research on this subject.
\section{Exact pinned curved solitary wave solution}
In the previous section, we have found a family of exact pinned accelerated  solitary wave
solutions, the velocity of which changes over time instead of being constant. Such accelerated
solitary waves move simultaneously with the charged debris object; hence are called pinned
solitary waves. The amplitude of dust ion acoustic solitary wave remains constant whereas both
the amplitude and velocity of debris function changes causing shape changing effects.
Generally, in real space plasma environment, the density localization may not be always of the
form of line solitons. It may bend, twist or turn during its propagation. Hence, to model such real
physical phenomena, solitary wave solutions with arbitrary free functions may be useful that
have such bending properties. Hence, along with arbitrary accelerated solitary waves, curved
solitary waves on $x-y$ plane are also important. In this section, we will try to find such kind of
exact solitary wave solutions which can bend on $x-y$ plane depending on the forcing debris
function. In \cite{SenKumar}, a numerical simulation for the propagation of magnetosonic wave
is performed with a two dimensional circular source term.
A noticeable difference from 1D simulation is observed in the shapes of wakes and precursors;
which are curved in nature. This observation also strengthens the possibility of existence of
curved solitary waves in experiment. There is no generalized exact solution of forced KP equation. In most of the cases, the forcing function
destroys the complete integrability of the equation; hence its exact solvability is also lost. But for certain special localized forcing functions, we can get exact solutions.
\begin{figure}
\includegraphics[width=8cm]{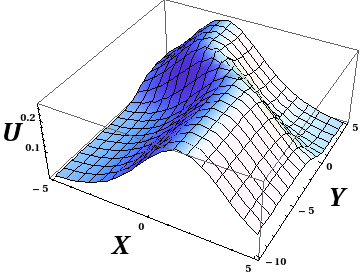}

\vspace{1cm}


\vspace{1cm}

\includegraphics[width=8cm]{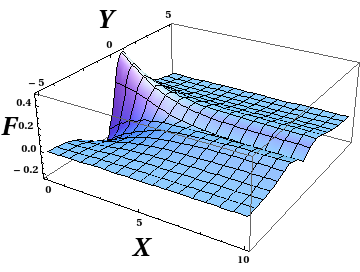}

\vspace{1cm}


\vspace{1cm}

\caption{The above figures show 3D plots of DIAW solution $U$ and debris function $F$ given by equations (\ref{csol}) and (\ref{Fcurv}) respectively on $X-Y$ plane at $T=0$, $c_1=0.5$ and $\theta_0 = 0$ for the choice of arbitrary function as : $A(Y) = \int sech{Y} dY$. It can be easily seen that both
$U$ and $F$ get curved on $X-Y$ plane.} \label{Fig4}
\end{figure}
 An exact pinned solitary wave solution $U$ of (\ref{KPnorm}) that can bend on $X-Y$ plane is
obtained in the following form
\begin{equation}
U = \frac{c_1}{2} \ Sech^2[\frac{\sqrt{c_1}}{2} \ (X + A(Y) - c_1 T + \theta_0)],\label{csol}
\end{equation}
for the choice of the forcing debris function $F$ as
\begin{align}
&{}F = A_Y^2 \ \frac{c_1}{2} \ Sech^2[\frac{\sqrt{c_1}}{2} \ (X + A(Y) - c_1 T + \theta_0)] +
\sqrt{c_1}\ A_{YY} ( \ \ \tanh[\frac{\sqrt{c_1}}{2} \ (X + A(Y) - c_1 T + \theta_0)] - \ 1), \ X > 0,
\nonumber \\
&{}F = A_Y^2 \ \frac{c_1}{2} \ Sech^2[\frac{\sqrt{c_1}}{2} \ (X + A(Y) - c_1 T + \theta_0)] +
\sqrt{c_1} \ A_{YY} ( \ \ \tanh[\frac{\sqrt{c_1}}{2} \ (X + A(Y) - c_1 T + \theta_0)] + \ 1) , \ X\leq 0
\label{Fcurv}
\end{align}
where $c_1$ is a constant and $A(Y)$ is an arbitrary function of Y. The function $F$ in Eq.
(\ref{Fcurv}) vanishes as $X \longrightarrow \pm \infty$ satisfying Poisson's equation
(\ref{Poisson}) as discussed before. There might be other choices of $F$ to get exact curved
solitary wave solution $U$ of (\ref{KPnorm}), but we consider (\ref{Fcurv}) as the first example of
$F$ to explore the situation.
The nonlinear function $A(Y)$ makes the solitary wave to bend in $X-Y$ plane. Both the
solutions $U$ and $F$ given by equation (\ref{csol}) and (\ref{Fcurv}) are pinned solutions, like
the solutions discussed in previous section. As we have shown in the previous section, the
pinned nature can be understood by plotting both $U, F$ with $X$ at different times. The 3D
plots of both $U$ and $F$ given by the above equations (\ref{csol}) and (\ref{Fcurv}) for a given
choice of $A(Y)$ are provided in Figure-\ref{Fig4}.
Thus we have provided an exact pinned curved solution $U$ for Eq. (\ref{KPnorm}) for a
specific choice of the localized function $F$. The solution may be used for modeling the twist or
turn of the solitary waves during its motion in real experiments. Though there may be more
general numerical or approximate solutions, but the exact form of this curved solution 
makes it important in this field.

\section{Results and Discussions}
 A few specific points of this work, that need to be discussed,  are stated below.
\begin{enumerate}
\item
In this work, we have derived few (2+1) dimensional exact dust ion acoustic
solitary wave solutions for choice of specific localized debris functions. The constant amplitude
exact solitary waves $U$ may accelerate or bend on $X-T$ and $X-Y$ planes respectively due
to specific choices of the forcing functions as discussed in section-III and section-IV.
\item
We can see that the coefficient of each term in equation (\ref{KP_original}) depends on $v_p$
which depends on $\alpha$ via equation (\ref{vp}), i.e,
\begin{equation}
v_p = \frac{1}{\sqrt{1-\alpha}}, \ \alpha = Z_d \frac{n_{d0}}{n_{i0}}.
\end{equation}
Hence, we can see that $\alpha $ must be $< 1$ for a real $v_p$.
The variation of phase velocity $v_p$ with parameter $\alpha$ is plotted in Figure-\ref{Fig6},
where we see that $v_p$ increases with $\alpha$. For ion acoustic wave, we can evaluate
$v_p$ to be = 1. The static dust grains increase the phase velocity by a factor $
\frac{1}{\sqrt{1-\alpha}}.$
\begin{figure}[hbt!]
\centering
\includegraphics[width=8cm]{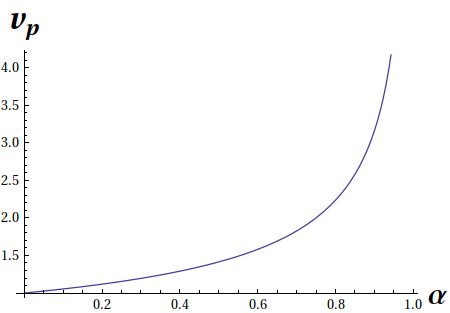}
\vspace{1cm}
\caption{Variation of phase velocity $v_p$ with $\alpha$.}\label{Fig6}
\end{figure}
\item
The solitary wave gets accelerated / decelerated due to presence of nonlinear function $a(T)$.
Obviously, for $a(T) =$ constant, we get the standard line solitary waves having constant
velocity. The acceleration of the solitary wave is shown by the contour plots in Figure-\ref{Fig7}
and Figure-\ref{Fig8}.
\begin{figure}[hbt!]
\centering
\includegraphics[width=8cm]{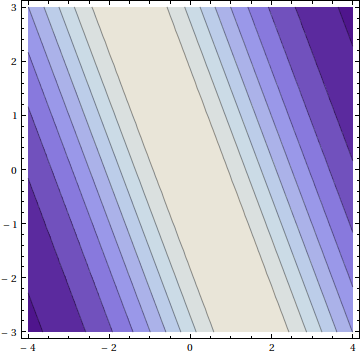}
\vspace{1cm}
\caption{The contour plot of the DIAW solution $U$ (\ref{solpinA1}) in $X-T$ plane for $Y=0$,
$a(T) = 2, d_1=1.5$. We can see that the peak of the solitary wave $U$ moves along a straight
line in $X-T$ plane for $a(T)=$ constant, thus showing the usual line solitary waves.}\label{Fig7}
\end{figure}
\begin{figure}[hbt!]
\centering
\includegraphics[width=8cm]{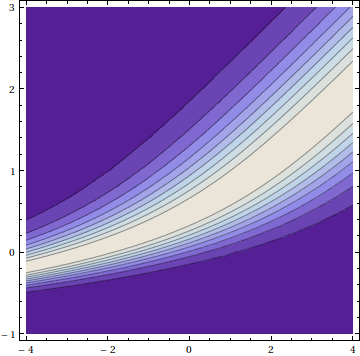}
\vspace{1cm}
\caption{The contour plot of the DIAW solution $U$ (\ref{solpinA1}) in $X-T$ plane for $Y=0$,
$a(T) = -5 e^{-2T}, d_1=2$. We can see that the peak of the solitary wave $U$ moves along a
curved path in $X-T$ plane for the exponential form of $a(T)$.}\label{Fig8}
\end{figure}
\item
The exact accelerated solitary wave solution (\ref{solpinA1}), which we have found, can be
expressed in old variables as
\begin{equation}
n_1 = \frac{3B(d_1-1)}{A} \ sech^2[\frac{\sqrt{d_1-1}}{2} \{ \xi + \sqrt{(B/C)} \eta - d_1 B \tau -
(\frac{60B^3}{DA}) \int \exp{(-2B \tau)} d\tau \}], \label{AccOld}
\end{equation}
where the coefficients $A, B, C$ are given as
\begin{eqnarray}
&{}A = \frac{v_p (3 - v_p^2)}{2}, \ B = \frac{v_p^3}{2}, \ C = D = \frac{v_p}{2}, 
\end{eqnarray}
for $a(T) = -5e^{-2T}$.
The variation of the DIAW $n_1$ in (\ref{AccOld}) with $\alpha$ at the center $\xi = \eta = 0$ at
$\tau =1$ is shown in Figure-\ref{Fig9}. 
\begin{figure}[hbt!]
\centering
\includegraphics[width=8cm]{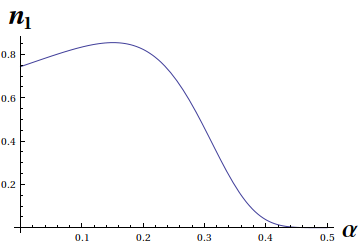}
\caption{ The variation of the DIAW $n_1$ in (\ref{AccOld}) with $\alpha$ at the center $\xi = \eta
= 0$ at $\tau =1$ for $d_1$ = 1.5.}\label{Fig9}
\end{figure}
\item
In this work, we have also derived an exact curved solitary wave solution for a special localized
debris function. The exact solitary wave solution can bend on $x-y$ plane depending on the
forcing function. In \cite{SenKumar}, a two
dimensional circular source term is considered in their numerical simulation for the propagation
of magnetosonic wave. The noticeable difference from the 1D simulation is in
the shapes of wakes and precursors which are curved in nature. This observation also
strengthens the possibility of observing the exact curved solitary wave solution (\ref{csol}) in
experimental scenario. It should be noted that curvature of the solution (\ref{csol}) occurs due
to presence of nonlinear function $A(Y)$ in the argument of the solution (\ref{csol}). Now it is
required to determine how much curvature is taking place by varying $A(Y)$ i.e, what the
condition is for larger
bending. For static case ($T = 0$), the locus of the maximum amplitude of the curved solitary
wave solution (\ref{csol}) is of the form:
\begin{equation}
\frac{\sqrt{c_1}}{2} \ (X + A(Y) + \theta_0) = 0. \label{locus}
\end{equation}
Now taking derivative w.r.to $Y$ twice in the above equation (\ref{locus}) we get,
\begin{equation}
\frac{d S}{d Y} = - A_{YY}, \label{condbend}
\end{equation}
where the slope is defined as: $S = \frac{d X}{d Y}.$
We see from (\ref{condbend}) that for higher value of RHS, rate of variation of the slope $S$
will be higher. Hence in that case,
the slope of the maximum amplitude curve will vary large for traversing unit distance in $Y$.
Larger rate of variation
of slope $S$ describes larger bending.
Hence, for large bending of solitary waves to take place, the double derivative of $A(Y)$ must
also be high.
This is explained clearly by the contour plot in Figure-\ref{Fig10}.
\begin{figure}
\includegraphics[width=8cm]{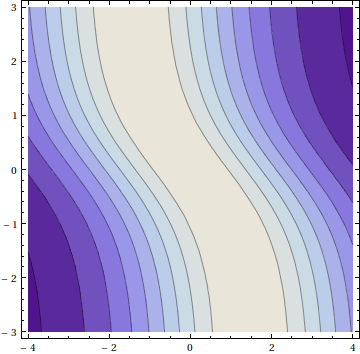}
\caption{Contour plot of curved dust ion acoustic solitary wave solution $U$ given by equation
(\ref{csol}) on $X-Y$ plane at $T=0$ for $c_1=0.5$, $\theta_0 = 0$ and $A(Y) = \int sech{Y}
dY$. The figure shows that DIAW solution $U$ bends on $X-Y$ plane due to the function
$A(Y)$.}\label{Fig10}
\end{figure}
\item
{Similar to the exact solutions obtained in \cite{Sen}, we derive few intricate exact solitary wave
solutions. The only difference is the solutions derived in \cite{Sen} are line solitary waves
whereas solutions derived in the present work are accelerated and curved solitary waves.
Basically we have investigated few specific forms of $F$ for which we get exact solitary wave
solutions for $U$ having bending features. The velocity profiles for both $U$ and $F$ are
obtained to be identical. For other general choice of $F$, perturbative or numerical solutions
need to be obtained}
\end{enumerate}
\section{Concluding remarks}
In this work, we have considered the low temperature and low density plasma in the LEO
region in presence of localized charged space debris particles. The dynamics of (2+1)
dimensional nonlinear dust ion acoustic waves induced in the system is found to be governed
by a forced KP-II equation, where the forcing term depends on the charged space debris
function. We have found some exact curved solitary wave solutions for the DIAW that can bend
on $x-t$ and $x-y$ planes respectively.  A family of exact pinned accelerated solitary wave solutions (\ref{debris2}) - (\ref{Ndeb}) have been derived. The velocity of the solutions changes over time whereas the amplitude remains constant. The solutions contain an arbitrary time dependent function $a(T)$ that can be chosen accordingly for modeling different types of dynamics of the solutions.
Also, a
special exact solitary wave solution (\ref{csol}) has been derived for a special debris function (\ref{Fcurv}) that gets curved on $x-y$ plane; where its
curvature depends on the nature of the forcing debris function. The solution also contain an arbitrary function $A(Y)$. For different choices of $A(Y),$ we would get different kinds of bending of the solitary wave. Thus, the exact solutions of this work  may be interesting  to the nonlinear
dynamics community and useful for  different practical applications.
\section{acknowledgements}
Siba Prasad Acharya acknowledges Department of Atomic Energy (DAE) of Government of
India for financial help during this work through institute fellowship scheme. Abhik Mukherjee
would like to acknowledge Indian Statistical Institute, Kolkata for the financial support during the
research work. Abhik Mukherjee is indebted to Ms. Krishna Kar for the scintillating discussions
and unconditional support during the progress of the work. The authors are immensely grateful
to the anonymous reviewers for their fruitful suggestions to improvise this work.

\end{document}